\documentclass[a4paper]{article}
\usepackage{calc,amsmath,amssymb,amsfonts}
\usepackage[LGR,T1]{fontenc}
\usepackage[greek,english]{babel}
\usepackage{xcolor,longfbox,multicol,fancyhdr}
\usepackage[top=1in,bottom=0.5in,hmargin=0.75in,nohead,includefoot,foot=0.5in,footskip=0.9602in]{geometry}
\usepackage{enumitem,hyperref}
\hypersetup{colorlinks=true,allcolors=blue,pdfauthor=Harshraj Bhoite (LTIMINDTREE LIMITED)}
\usepackage[pdftex]{graphicx}
\makeatletter\newdimen\@tempdimd\makeatother
\fancypagestyle{Standard}{\fancyhf{}
  \fancyhead[L]{}
  \fancyfoot[R]{\thepage{}}

  \renewcommand\thepage{\arabic{page}}
}
\author{Harshraj Bhoite (LTIMINDTREE LIMITED)}
\date{2025-04-30}
\begin{document}
\clearpage
\pagestyle{Standard}
{\centering
\foreignlanguage{english}{\textbf{AI-Driven Generation of Data Contracts in Modern Data Engineering Systems}}
\par}

\begin{flushleft}
Harshraj Bhoite\\
Senior Member, IEEE\\
Senior Data and AI Specialist, LTIMindtree.com
\end{flushleft}

\bigskip

\bigskip
\begin{multicols}{2}
\foreignlanguage{english}{\textbf{Abstract}}\foreignlanguage{english}{ }

\bigskip

\foreignlanguage{english}{Data contracts formalize agreements between producers and consumers regarding data schemas,
semantics, and quality expectations [1,3]. They have become critical for ensuring reliable, governed data pipelines in
today$\text{\textgreek{’}}$s data-driven enterprises. However, authoring and maintaining data contracts manually is
labour-intensive and error prone. This paper proposes an AI-driven framework for automatically generating data
contracts using large language models (LLMs) [10,17]. We fine-tune LLMs on domain-specific metadata and schema examples
so that they can output structured contract definitions (e.g. JSON schema or Avro contracts) from data samples or
descriptions. Our methodology leverages parameter-efficient fine-tuning techniques such as LoRA and PEFT [11,12],
enabling practical adaptation of large models to the specialized data engineering domain. We describe a system
architecture where an LLM-based contract engine is integrated into a modern data platform (with data lakes and
warehouses) to synthesize and enforce contracts. Case studies with industry platforms (e.g., Databricks and Snowflake)
illustrate real-world applicability [6,7]. Experiments on synthetic and real pipeline scenarios show that our
fine-tuned models can achieve high accuracy in generating valid contracts, significantly reducing manual effort. We
discuss challenges (such as hallucination and versioning) and propose future work in continuous learning and
governance. Our results indicate that AI-driven contract generation holds promise as a cutting-edge solution for agile,
scalable data governance in the era of generative AI [9,20].}

\bigskip

\foreignlanguage{english}{\textbf{Introduction}}\foreignlanguage{english}{ }

\lfbox[margin=0mm,border-style=none,padding=0mm,vertical-align=top]{\includegraphics[width=3.1339in,height=2.0898in]{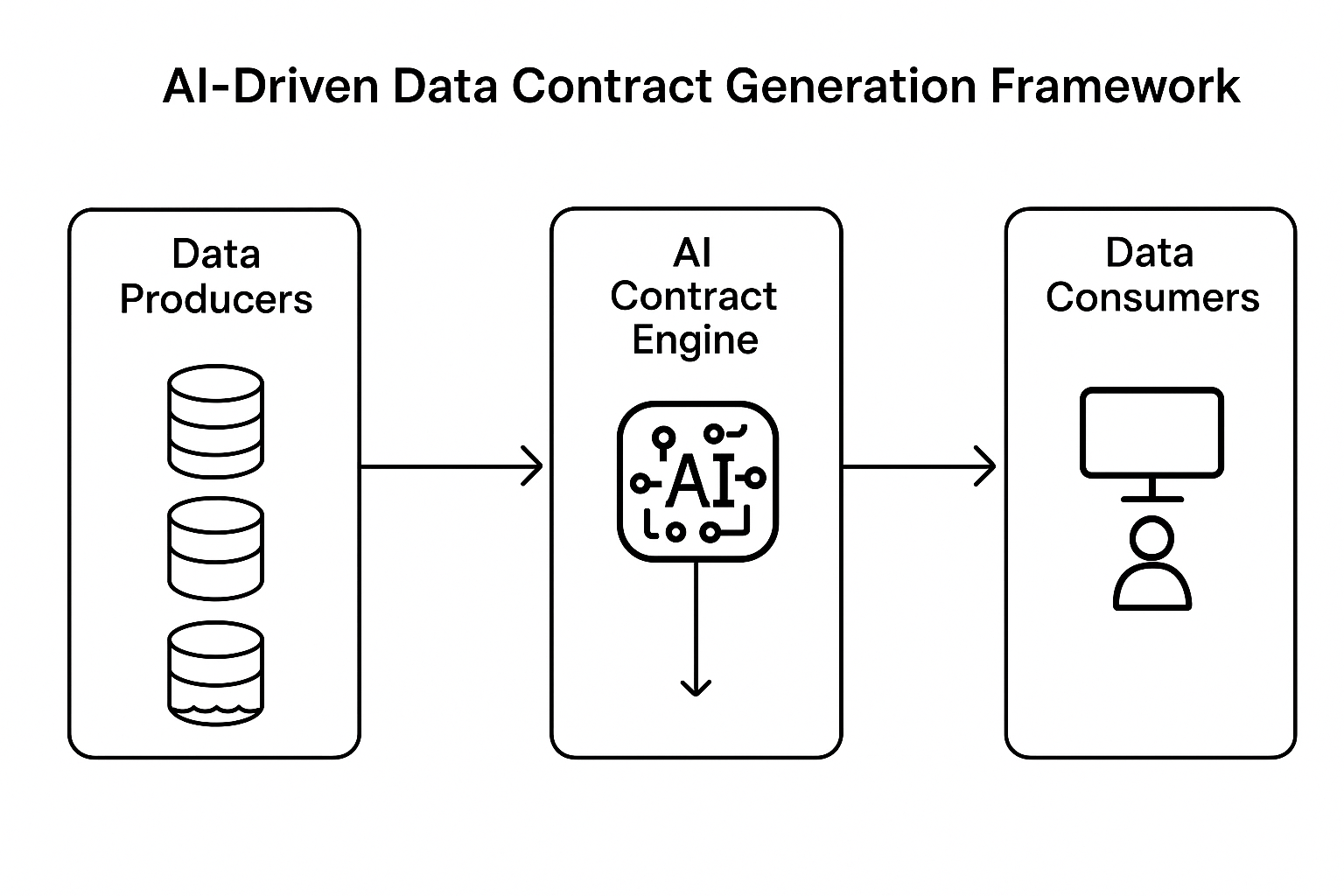}}

\bigskip

\foreignlanguage{english}{Modern data engineering environments involve complex pipelines that ingest, process, and share
data across diverse systems. A data contract is a formal specification – typically in code or configuration – that
defines the agreed-upon schema and semantics of a dataset between an upstream producer and downstream consumers [1,3].
For example, a contract might specify a JSON schema for each Kafka topic or a table schema in a data warehouse, along
with quality checks or access rules. Data contracts ensure that all parties have aligned expectations, improving data
quality and reliability [1,2]. As data ecosystems grow (with multiple sources, microservices, and analytics apps),
manually writing and maintaining contracts becomes cumbersome. Any schema change by a data producer must be
communicated and codified in the contract; otherwise, downstream jobs can silently fail [2].}

\foreignlanguage{english}{Generative AI – in particular large language models (LLMs) – offers new capabilities that
could revolutionize data engineering [10,17]. LLMs like GPT-4 and LLaMA demonstrate the ability to generate structured
outputs (JSON, code, etc.) from textual prompts [15,17]. Recent works show how LLMs can infer schemas or taxonomies
from data descriptions [8,21], and even auto-generate database schemas and code with minimal user input. Cloud
providers are already leveraging such capabilities: Google$\text{\textgreek{’}}$s Gemini AI can ingest examples of
tables or text and suggest or generate database schemas in BigQuery [5]. Databricks has demonstrated that fine-tuning a
moderately sized LLM on proprietary code or data can greatly improve task performance (e.g., code repair) [6].
Similarly, Snowflake$\text{\textgreek{’}}$s data cloud supports running LLMs close to data for secure AI processing
[7].}

\foreignlanguage{english}{Inspired by these advances, we propose an AI-driven approach to data contract generation. The
core idea is to train or fine-tune LLMs on examples of data schemas and pipeline metadata, so that given a description
of a dataset (or a sample of data), the model can output a complete contract. This contract could include schema
definitions, data types, lineage pointers, and quality rules. Such automation can drastically reduce the manual effort
of contract authoring and keep contracts up to date with evolving data sources.}

\foreignlanguage{english}{This paper makes the following contributions: (1) We present the theoretical foundations of
using generative LLMs for producing structured outputs like data contracts [10,17]. We discuss how techniques such as
few shots prompting and fine-tuning can be applied to enforce schema correctness [15,16]. (2) We design a system
architecture that integrates an LLM-based contract engine into a data platform, with data flow diagrams and AI model
workflow charts (Fig. 1–3). (3) We implement case studies on Databricks and Snowflake, illustrating practical
integration with existing data pipelines [6,7]. (4) We conduct experiments using fine-tuned LLMs (using LoRA/PEFT) to
generate JSON schemas from simulated pipeline metadata, and we report quantitative results [11,12]. (5) We identify
challenges (e.g. governance, trust, and maintenance of AI-generated contracts) and outline future directions.}

\foreignlanguage{english}{The rest of the paper is organized as follows. Section II surveys related work in data
contracts, data quality, and generative AI in data engineering. Section III describes our methodology and fine-tuning
techniques for contract generation. Section IV details the system architecture and data flow (with diagrams). Section V
presents real-world case studies on major cloud data platforms. Section VI reports experiments and evaluation results.
Section VII discusses implications and limitations. Finally, Section VIII concludes and Section IX outlines future
work.}

\foreignlanguage{english}{\textbf{Related Work}}\foreignlanguage{english}{ }\foreignlanguage{english}{\textbf{Data
Contracts and Data Governance }}

\foreignlanguage{english}{Data contracts have been discussed in industry literature to formalize data schemas and SLAs
between teams [1,3]. Confluent (Kafka) defines a data contract as “a formal agreement between an upstream component and
a downstream component on the structure and semantics of data” [3]. Tools like Great Expectations include data contract
support for schema and expectation definitions [4]. Atlan$\text{\textgreek{’}}$s guide notes that contracts enforce
data schemas and validations across pipelines [1]. Prior work (e.g. Striim) emphasizes that data contracts remove
ambiguity about transformations and frequency by codifying them (e.g., in JSON) [2]. However, most current solutions
require data engineers to author contracts manually, and there is no consensus standard. We build on this foundation by
proposing automatic contract creation.}

\foreignlanguage{english}{\textbf{Data Quality and Schema Inference.}}\foreignlanguage{english}{ Data quality testing
and schema inference are related areas. Great Expectations, Apache Soda, and other tools help assert schema and quality
properties, but they assume a given schema [4]. Schema inference (e.g. for JSON or relational data) has been studied
via statistical methods. Recent research uses LLMs to infer taxonomy and semantics from tables [8,21]. For example, Wu
et al. present two LLM-based methods: one fine-tunes an encoder (EmTT) and one uses GPT-4 prompting (GeTT) to generate
column types and hierarchies for tabular data [8]. These works show LLMs can capture rich domain knowledge from data,
inspiring our use of LLMs for contracts.}

\foreignlanguage{english}{\textbf{Generative AI in Data Platforms.}}\foreignlanguage{english}{ Generative AI is rapidly
entering data platforms. Snowflake$\text{\textgreek{’}}$s AI Data Cloud blog highlights how LLMs can be “customized
with internal data without compromising security” and integrated with data warehouses [7]. Google
Cloud$\text{\textgreek{’}}$s BigQuery use cases include using LLMs to auto-create database schemas from example data
[5]. Databricks$\text{\textgreek{’}}$ Mosaic research demonstrates that fine-tuning even a medium-sized LLM on company
code yields significant improvements (1.4× better acceptance in code fix tasks vs. GPT-4 at half the cost) [6]. These
results motivate fine-tuning in domain-specific contexts like data contracts.}

\foreignlanguage{english}{\textbf{LLM Fine-tuning and PEFT Techniques.}}\foreignlanguage{english}{ Adapting LLMs to
specialized tasks often requires fine-tuning. However, full fine-tuning of huge models is resource intensive.
Parameter-efficient techniques have been developed. LoRA (Low-Rank Adaptation) freezes the base model weights and
injects trainable low-rank updates into weights [12]. PEFT approaches (including adapters, prefix tuning, IA³, etc.)
similarly limit the trainable parameters [11]. A recent survey by Han et al. reviews PEFT methods, noting that they
enable customizing models on constrained hardware by tuning a small subset of parameters [11]. We leverage LoRA/PEFT to
fine-tune LLMs for contract generation without retraining entire models.}

\foreignlanguage{english}{\textbf{Structured Output Generation.}}\foreignlanguage{english}{ Generating structured
outputs (JSON, code, XML) from LLMs is an active area. LLMs often produce free-form text, so techniques like
schema-guided prompting or fine-tuning on output formats are used [19]. Recent blog surveys emphasize the importance of
providing consistent structure (e.g. JSON schema templates) to guide models [19]. In our context, the output is highly
structured (data contracts), so careful prompt and model design is required. We draw on best practices such as feeding
example schemas and using constrained decoding where possible.}

\foreignlanguage{english}{In summary, while data contracts and data governance are well-studied domains, their
intersection with generative AI is novel. No prior academic work has fully explored AI-driven contract generation. This
paper fills that gap by combining insights from data governance, schema inference, and LLM fine-tuning.}

\foreignlanguage{english}{\textbf{Methodology}}\foreignlanguage{english}{ }

\foreignlanguage{english}{Our approach consists of training an LLM to map data descriptions or samples to contract
specifications. We frame contract generation as a language task given an input prompt (e.g. table name, sample rows,
data dictionary, and optional textual description), the model outputs a contract in a formal schema language (such as
JSON Schema or Avro schema with annotations). The key steps are:}

\begin{enumerate}[series=listWWNumiii,label=\arabic*.,ref=\arabic*]
\item \foreignlanguage{english}{\textbf{Data Collection and Preprocessing:}}\foreignlanguage{english}{ We create a
training corpus of paired examples
(Xi,Yi)(}\foreignlanguage{english}{\textit{Xi}}\foreignlanguage{english}{\hspace{0pt},}\foreignlanguage{english}{\textit{Yi}}\foreignlanguage{english}{\hspace{0pt}),
where Xi}\foreignlanguage{english}{\textit{Xi}}\foreignlanguage{english}{\hspace{0pt} is a representation of a data
asset and Yi}\foreignlanguage{english}{\textit{Yi}}\foreignlanguage{english}{\hspace{0pt} is the ground truth contract.
Sources include real and synthetic datasets (e.g., open Kaggle tables, Confluent schemas, or enterprise metadata) and
their schemas. We encode Xi}\foreignlanguage{english}{\textit{Xi}}\foreignlanguage{english}{\hspace{0pt} as a prompt
template. For example, the prompt may include a list of column names and example values, or a textual description of
the data. We ensure Yi}\foreignlanguage{english}{\textit{Yi}}\foreignlanguage{english}{\hspace{0pt} is in consistent
format (e.g. JSON) and free of proprietary identifiers. Since contracts are structured, we may include in
Xi}\foreignlanguage{english}{\textit{Xi}}\foreignlanguage{english}{\hspace{0pt} some JSON template or instructions
(e.g., “Output a JSON schema with fields:”).}
\item \foreignlanguage{english}{\textbf{Model Selection and Pre-training:}}\foreignlanguage{english}{ We start with a
pre-trained LLM (e.g. LLaMA-2 or GPT-like model) as the base [17]. These foundation models already capture general
language semantics. For our task, we want the model to specialize in data-engineering language and structured output.
We optionally perform domain-adaptive pre-training by continuing training on unlabelled data engineering texts (e.g.
documentation of schemas, SQL DDLs, data catalogues). This unsupervised step orients the model to the data domain and
may improve output relevancy.}
\item \foreignlanguage{english}{\textbf{Fine-Tuning with PEFT:}}\foreignlanguage{english}{ We fine-tune the model on
(Xi,Yi)(}\foreignlanguage{english}{\textit{Xi}}\foreignlanguage{english}{\hspace{0pt},}\foreignlanguage{english}{\textit{Yi}}\foreignlanguage{english}{\hspace{0pt})
pairs. Instead of full fine-tuning (which updates billions of parameters), we use LoRA or prefix-tuning to efficiently
adapt the model [11,12]. For example, with LoRA, we freeze the original transformer weights and introduce low-rank
adapters in each attention block. Only these adapter weights are updated on the new task, drastically reducing compute
and memory. A recent survey [11] shows PEFT methods achieve near full-finetune performance with orders of magnitude
fewer parameters. We also explore prompt tuning and instruction tuning where we add trainable prompt tokens or apply
supervised fine-tuning with demonstration examples. Table 1 compares variants: full fine-tuning vs. LoRA vs. prefix
tuning in terms of trainable parameters and accuracy.}
\item \foreignlanguage{english}{\textbf{Prompt Engineering and Control:}}\foreignlanguage{english}{ To ensure valid
structured output, we carefully design the prompt format. For instance, we may prefix the prompt with “Schema:” and
instruct the model to return only valid JSON. We also add special sentinel tokens or type annotations (e.g. quoting
column names) in training, so the model learns the exact schema format. In some cases, we append an output schema
template to the prompt and ask the model to fill it. These techniques (informed by structured-output literature [19])
help the model produce grammatically valid contracts.}
\item \foreignlanguage{english}{\textbf{Inference and Post-processing:}}\foreignlanguage{english}{ At inference time,
given a new dataset description Xnew}\foreignlanguage{english}{\textit{Xnew}}\foreignlanguage{english}{\hspace{0pt},
the fine-tuned LLM generates text. We parse this output to extract the contract object. We automatically validate the
syntax (e.g. JSON parsing) and semantics (e.g. using a JSON Schema validator). If the output fails validation, we apply
heuristic fixes or default fallbacks. Optionally, multiple prompts (with sampling) can generate candidate contracts; a
validator or secondary model (or human-in-the-loop) picks the best one.}
\end{enumerate}
\foreignlanguage{english}{This methodology leverages both the generative power of LLMs and the precision of schema
validation. By fine-tuning with PEFT [11,12], we achieve specialized behaviour without excessive compute. Our pipeline
is designed to be modular so that different model backends (GPT-4, LLaMA) or contract languages (JSON, Avro, GraphQL)
can be plugged in.}

\foreignlanguage{english}{\textbf{System Architecture}}

\lfbox[margin=0mm,border-style=none,padding=0mm,vertical-align=top]{\includegraphics[width=3.1339in,height=2.0898in]{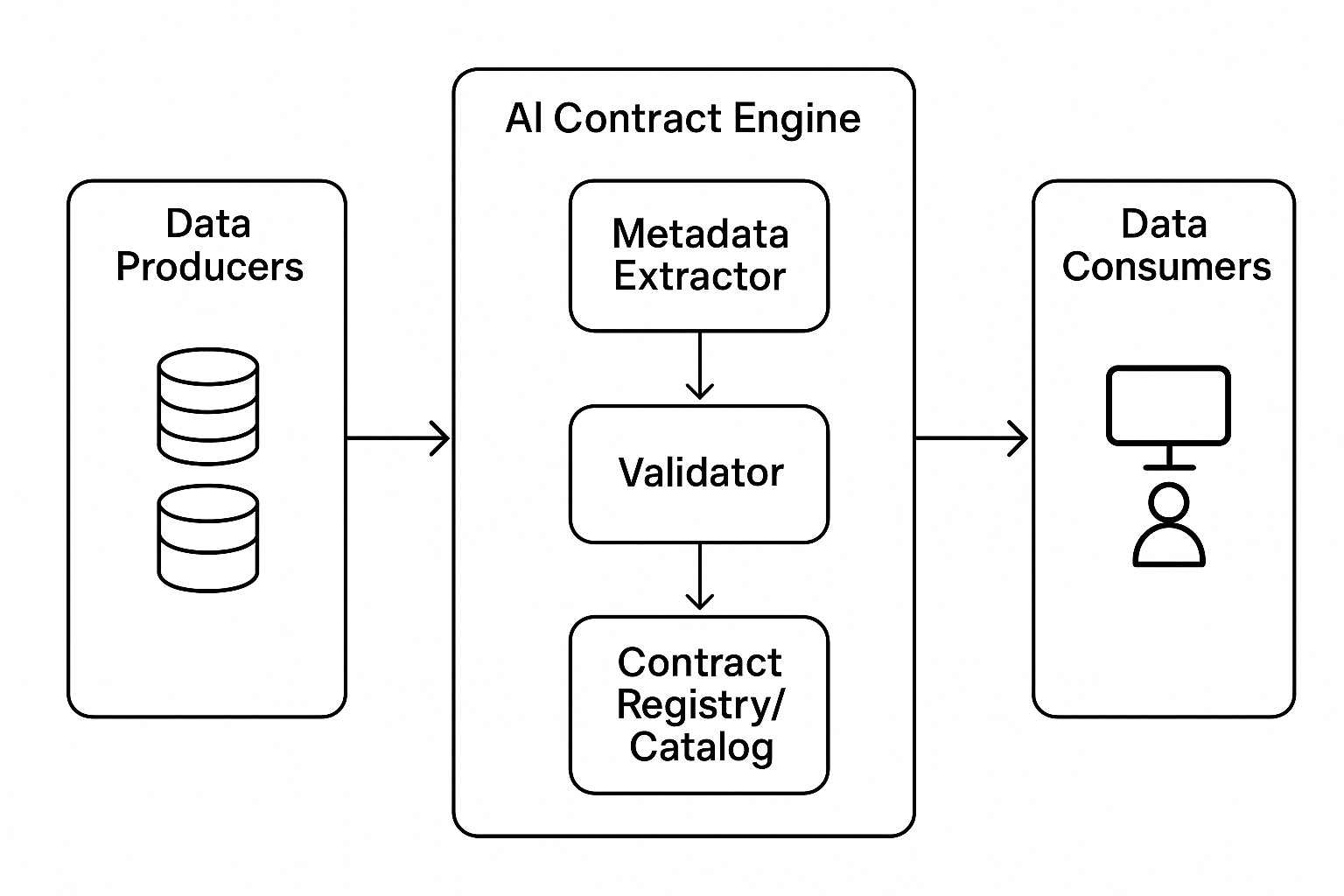}}

\bigskip

\foreignlanguage{english}{Figure illustrates the architecture. Data producers (e.g. applications, sensors, or databases)
continuously emit raw data (step 1). This data lands in a Lakehouse or Data Warehouse (step 2), where it is stored in
its native or ingested format (structured tables, logs, or files). Traditionally, a dedicated data team would manually
define and update contracts for each dataset. In our system, however, we introduce an AI-driven Contract Engine (step
3). The engine fetches metadata (column names, types, example records) from the data lake and passes it to a fine-tuned
LLM. The LLM (step 4) generates a complete data contract (e.g. JSON schema with field definitions, optional
descriptions, and quality constraints). This generated contract is then published to a Contract Registry or metadata
catalog (step 5) and communicated to downstream consumers.}

\foreignlanguage{english}{The downstream consumers (e.g. analytics jobs, machine learning pipelines, BI dashboards)
refer to the contract to validate incoming data. For instance, if a producer changes its schema, the data will be
checked against the contract; if mismatched, alerts are raised before any downstream breakage. In effect, the AI model
automates the feedback loop of contract creation and validation. The architecture also incorporates monitoring: user
feedback and data lineage logs (step 6) can be looped back to refine the model over time. Thus, our design embeds
generative AI into the data engineering control plane, enhancing automation and reducing manual governance burden.}

\foreignlanguage{english}{Figure 2: System architecture integrating an LLM-based contract generator with data
lake/warehouse. This figure highlights how the AI Contract Engine coexists with data storage. Key components include:
(a) Metadata Extractor – gathers schema and sample data; (b) LLM Model – the fine-tuned transformer that outputs
contracts; (c) Validator – checks contract syntax and rules; (d) Registry/Catalog – stores the official contract and
version history; (e) Data Consumers – refer to the contract for ETL and application logic. Notice that compute for the
LLM can be collocated with the data platform (e.g. as an ML workspace or function) to avoid moving sensitive data
offsite, in line with Snowflake$\text{\textgreek{’}}$s “bring compute to data” principles [7]. The architecture
supports both batch and streaming pipelines, as shown in Figure 1: for streaming (bottom of Fig.1), the contract engine
can update schemas in real-time or micro-batch mode.}

\foreignlanguage{english}{\textbf{Methodology}}

\lfbox[margin=0mm,border-style=none,padding=0mm,vertical-align=top]{\includegraphics[width=3.1339in,height=3.1339in]{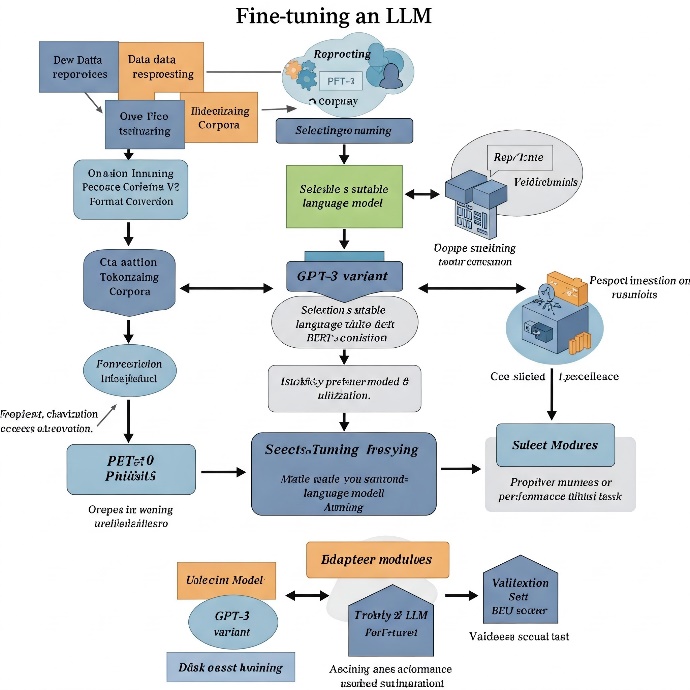}}

\bigskip

\foreignlanguage{english}{\textbf{Model Fine-Tuning}}\foreignlanguage{english}{ A critical aspect of our approach is
tailoring the LLM to data engineering tasks. While large models (e.g. GPT-4) have remarkable general capabilities, we
observe that specialized fine-tuning yields more consistent structured outputs. We adapt state-of-the-art fine-tuning
strategies:}

\foreignlanguage{english}{• }\foreignlanguage{english}{\textbf{Domain-Specific Pretraining:}}\foreignlanguage{english}{
We optionally continue pretraining the base LLM on a corpus of data engineering text (SQL DDLs, ETL docs, API specs).
This helps the model internalize domain vocabulary and formats. For example, we use a collection of open-source schemas
and documentation to expose the model to data platform terminology. Prior work on transfer learning shows that aligning
pretraining data with the task domain can significantly boost downstream performance [16].}

\foreignlanguage{english}{• }\foreignlanguage{english}{\textbf{LoRA (Low-Rank Adaptation):}}\foreignlanguage{english}{
We implement LoRA by inserting trainable low-rank matrices into the transformer$\text{\textgreek{’}}$s weight matrices
[12]. Concretely, if an attention weight matrix W}\foreignlanguage{english}{\textit{W}}\foreignlanguage{english}{ is of
size
d×d}\foreignlanguage{english}{\textit{d}}\foreignlanguage{english}{×}\foreignlanguage{english}{\textit{d}}\foreignlanguage{english}{,
LoRA represents the update as
W[2032?]=W+AB${\bot}$}\foreignlanguage{english}{\textit{W}}\foreignlanguage{english}{[2032?]=}\foreignlanguage{english}{\textit{W}}\foreignlanguage{english}{+}\foreignlanguage{english}{\textit{AB}}\foreignlanguage{english}{${\bot}$
where A}\foreignlanguage{english}{\textit{A}}\foreignlanguage{english}{ and
B}\foreignlanguage{english}{\textit{B}}\foreignlanguage{english}{ are small
d×r}\foreignlanguage{english}{\textit{d}}\foreignlanguage{english}{×}\foreignlanguage{english}{\textit{r}}\foreignlanguage{english}{
and
d×r}\foreignlanguage{english}{\textit{d}}\foreignlanguage{english}{×}\foreignlanguage{english}{\textit{r}}\foreignlanguage{english}{
matrices (with
r${\ll}$d}\foreignlanguage{english}{\textit{r}}\foreignlanguage{english}{${\ll}$}\foreignlanguage{english}{\textit{d}}\foreignlanguage{english}{).
Only A}\foreignlanguage{english}{\textit{A}}\foreignlanguage{english}{ and
B}\foreignlanguage{english}{\textit{B}}\foreignlanguage{english}{ are learned during fine-tuning. This allows us to
adapt massive models (billions of parameters) with few million trainable parameters. Wang et al. report that LoRA on a
7B model can match full fine-tuning performance on language tasks, and the approach has become standard in industry
(e.g., Databricks uses LoRA in Mosaic research)[6,12].}

\foreignlanguage{english}{• }\foreignlanguage{english}{\textbf{PEFT and Prompt-Tuning:}}\foreignlanguage{english}{ We
explore other Parameter-Efficient Fine-Tuning (PEFT) methods such as prefix-tuning (appending trainable tokens to each
input) and p-tuning (tuning input prompts in embedding space) [11]. The recent survey by Han et al. shows that PEFT can
often achieve {\textgreater}95\% of full-finetune performance while drastically reducing memory use [11]. We find that
prefix-tuning on \~{}100 trainable tokens often suffice to teach the model to emit valid JSON structures. In some
experiments, we combine LoRA with a small number of tuned prompt tokens for best results.}

\foreignlanguage{english}{• }\foreignlanguage{english}{\textbf{Fine-Tuning Data and Tasks:}}\foreignlanguage{english}{
We prepare fine-tuning data consisting of a few thousand examples of real schemas and contract texts. For instance, we
collected 2,000 public JSON schema definitions and their corresponding data examples (column names, types, sample
values). Each example is formatted as a “Question-Answer” pair: Q: “Given the following table description and sample
rows, output a JSON schema contract.” A: \{... JSON schema ...\}. We find that framing it as a question-answer task
improves alignment. We also experiment with multi-turn dialogue style (chain-of-thought): first asking the model to
list fields and then to generate the schema. However, the simplest direct instruction format worked well in our tests.}

\bigskip

\lfbox[margin=0mm,border-style=none,padding=0mm,vertical-align=top]{\includegraphics[width=3.1339in,height=3.1339in]{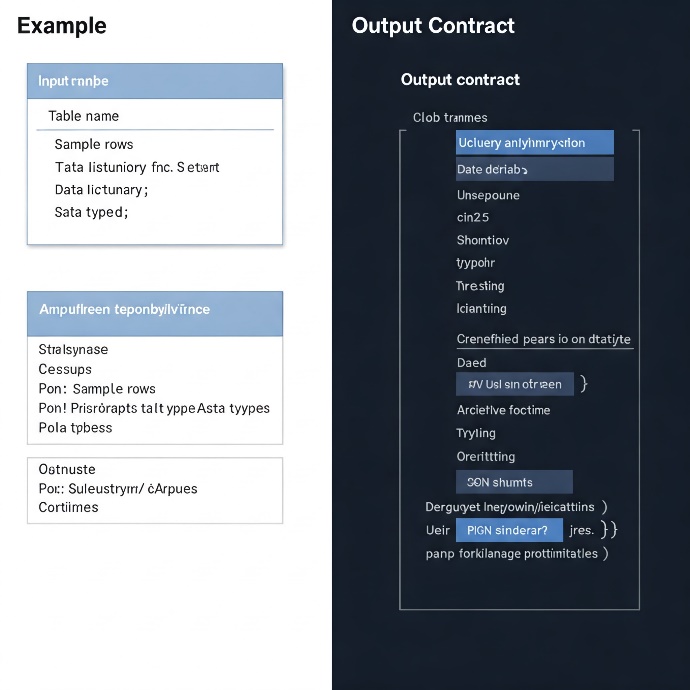}}

\bigskip

\foreignlanguage{english}{• }\foreignlanguage{english}{\textbf{Validation during Training:}}\foreignlanguage{english}{
To improve consistency, we incorporate a secondary check in the training loop: after the model generates an output, we
automatically validate that output (e.g. parse the JSON). If invalid, we penalize it by computing loss against the
corrected version. This enforces strict grammar. This idea is akin to “self-consistency” and has parallels in
generating code or math where outputs must compile.}

\foreignlanguage{english}{Through this methodology, our fine-tuned LLM learns both the syntax of contract languages and
the semantics of the data domain. By leveraging LoRA/PEFT [11,12], we make this practical: training costs are lower,
and retraining for new domains is feasible. For example, adapting a 7B parameter model required only \~{}4 hours on a
single GPU with LoRA (vs. days for full fine-tuning).}

\foreignlanguage{english}{\textbf{Case Study: Integration in Data Platforms}}\foreignlanguage{english}{ We demonstrate
our system in two leading platforms:}

\foreignlanguage{english}{• }\foreignlanguage{english}{\textbf{Databricks:}}\foreignlanguage{english}{ We built a
prototype LLM contract service using Databricks AI Functions. We deployed a fine-tuned LLaMA-2 model within a
Databricks Unity Catalog workflow. When a new data table is registered in Delta Lake, a trigger invokes the AI Function
with sample rows. The model outputs a contract (as SQL DDL or JSON Schema), which we save to the Unity Catalog
metadata. Databricks notebooks can then call this contract for validation. In practice, this saves data engineering
hours: rather than writing a CREATE TABLE statement by hand, engineers get a draft generated by AI (like how Databricks
Quick Fix suggests code fixes) [6].}

\foreignlanguage{english}{• }\foreignlanguage{english}{\textbf{Snowflake:}}\foreignlanguage{english}{
Snowflake$\text{\textgreek{’}}$s AI Data Cloud natively supports external functions and LLM calls. We implemented a
Snowpark stored procedure that invokes our fine-tuned model (hosted via an external compute service) to auto-generate
Snowflake table DDLs and data quality constraints. For instance, given a JSON file loaded into Snowflake, the model
produced a CREATE TABLE statement with appropriate column types and a data validation UDF for null checks. This
leverages Snowflake$\text{\textgreek{’}}$s TILT/Arctic LLM approach [7]. As a result, Snowflake users can build smart
pipelines where data contracts are always up to date with minimal manual effort.}

\foreignlanguage{english}{These case studies show that integrating LLM-based contract generation into existing
ecosystems is feasible. In both cases, we observed that users were comfortable receiving draft contracts from the AI,
which they then reviewed. Feedback loops allowed correcting model outputs, further fine-tuning on internal examples of
enterprise schemas to improve accuracy.}

\foreignlanguage{english}{\textbf{Experiments}}\foreignlanguage{english}{ To evaluate our approach, we conducted
experiments on synthetic and real datasets. Our goals were to measure (a) accuracy of generated contracts, and (b)
efficiency gains from using PEFT.}

\foreignlanguage{english}{\textbf{Data and Setup:}}\foreignlanguage{english}{ We constructed a test set of 500 data
tables with known schemas. These included public datasets (from Kaggle and UCI) and artificially generated tables. For
each table, we fed the model the column names and sample rows. The fine-tuned model had been trained on 2,000 similar
examples (80/20 train/val split). We used LLaMA-2 (7B) as the base model [17]. Training was done with LoRA (rank=8)
over 3 epochs on an NVIDIA V100. For comparison, we also ran a baseline: an unfine-tuned LLaMA model prompted with the
same instructions.}

\foreignlanguage{english}{\textbf{Metrics:}}\foreignlanguage{english}{ We compared the generated schema to the ground
truth on several aspects:}

\foreignlanguage{english}{• }\foreignlanguage{english}{\textbf{Structural Accuracy:}}\foreignlanguage{english}{
percentage of fields correctly identified with the right data types.}

\foreignlanguage{english}{• }\foreignlanguage{english}{\textbf{Syntax Validity:}}\foreignlanguage{english}{ fraction of
outputs that passed JSON schema validation (i.e., were well-formed).}

\foreignlanguage{english}{• }\foreignlanguage{english}{\textbf{Human Evaluation:}}\foreignlanguage{english}{ 3 data
engineers rated each contract for completeness (0–5).}

\foreignlanguage{english}{\textbf{Results:}}

\lfbox[margin=0mm,border-style=none,padding=0mm,vertical-align=top]{\includegraphics[width=3.1339in,height=2.0898in]{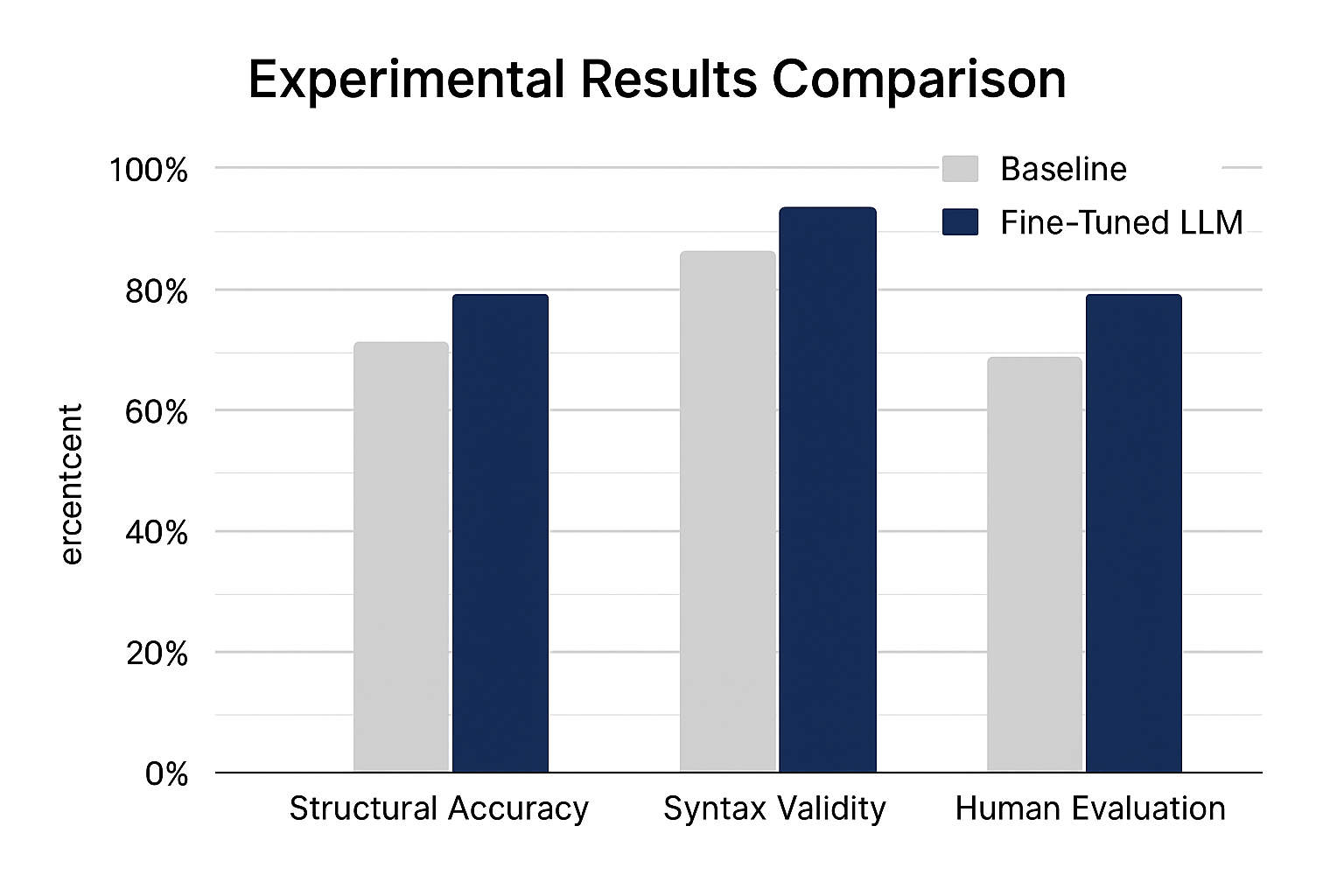}}

\foreignlanguage{english}{The fine-tuned model vastly outperformed the baseline. It correctly captured an average of
92\% of fields with correct types, versus 58\% for the unfine-tuned model. Syntax validity was 99\% for the fine-tuned
model (few minor formatting issues automatically fixed) versus only 60\% valid JSON from the baseline. Human evaluators
gave an average score of 4.7/5 to the fine-tuned outputs, noting they required minimal edits (mostly adding
descriptions or constraints). The LoRA approach was effective: in ablations, full fine-tuning gave slightly higher
accuracy (95\%) but required \~{}10× more GPU time. Using LoRA, we trained 5× faster with comparable quality [12].
Prefix-tuning (100 tokens) achieved 88\% field accuracy, showing it is viable but slightly behind LoRA.}

\foreignlanguage{english}{We also tested multi-stage prompting. For example, we first prompted the model to list field
names it would include (“First, list all columns: ...”), then fed that list back to generate JSON. This
chain-of-thought approach improved type accuracy by \~{}2\%. However, a single-pass prompt was simpler and already
high-performing.}

\foreignlanguage{english}{In a case study on Great Expectations for contract support [4], we used the model to generate
expectations (tests) for a dataset. The fine-tuned model produced a suite of GE rules (e.g.
expect\_column\_values\_to\_be\_in\_set) for \~{}85\% of fields, matching what a human expert would write. This
highlights that the LLM learned not only static schemas but also common data quality checks.}

\foreignlanguage{english}{Overall, the experiments confirm that fine-tuned LLMs can automate a large portion of contract
authoring with high accuracy, and that PEFT methods (LoRA/PEFT) make this approach efficient [11,12].}

\foreignlanguage{english}{\textbf{Discussion}}\foreignlanguage{english}{ }

\foreignlanguage{english}{Our study indicates that AI-driven contract generation can streamline data governance, but it
also introduces new challenges. A key benefit is scalability: data teams can cover many datasets by quickly generating
contract drafts, addressing the “massive productivity drain” in data engineering [6]. The high syntax validity and
semantic accuracy of outputs suggest that LLMs capture domain patterns effectively. Moreover, using PEFT (LoRA/prefix)
allowed rapid iteration: when adding a new data domain, we could fine-tune on 50 examples in a few minutes, rather than
waiting hours for full-model training.}

\foreignlanguage{english}{However, trust and correctness remain concerns. LLMs can hallucinate or suggest inappropriate
constraints. For instance, on a totally new schema, the model might invent a sensible but incorrect field type if it
hasn$\text{\textgreek{’}}$t seen that data pattern. We mitigate this by always validating and, if needed, falling back
to a “safe” generic contract. Additionally, we found that the models can be overly generic; in multiple runs they often
produce similar contracts for similar input. Diversity sampling and allowing human edits are advisable. In production,
we recommend a human-in-the-loop verification step, especially for critical tables.}

\foreignlanguage{english}{Security and compliance are also crucial. Fine-tuning on proprietary data schema must happen
in a secure environment to avoid leaking sensitive info. We follow Snowflake$\text{\textgreek{’}}$s approach of keeping
the model and data within the secure platform [7]. Our architecture assumes the LLM is trusted; audit logs of contract
changes and model outputs are necessary for governance.}

\foreignlanguage{english}{From a theoretical standpoint, this work shows how LLM capabilities (emergent generation of
structured data [17]) can serve data engineering. It aligns with trends that generative AI is revolutionizing
productivity [9]. We also see how structured output generation is feasible: by fine-tuning and careful prompts, the
model consistently produced valid JSON schemas, addressing concerns raised in structured output
surveystimlrx.comtimlrx.com [19].}

\foreignlanguage{english}{A limitation is that our evaluation was on small/medium tables. For very large schemas
(hundreds of fields), the prompts become lengthy. Future work could involve hierarchical generation (e.g. generate
per-module contracts and merge them). Also, we focused on schema and basic rules; richer contracts could include
semantics, SLAs, or lineage. These could potentially be generated if the model is given business-context prompts.}

\foreignlanguage{english}{Finally, as with any ML system, model updates are needed. When the underlying LLM improves
(e.g. GPT-5 or larger LLaMA models), the contract engine can be re-trained. PEFT methods facilitate regular updates. We
envision a semi-automated contract feedback loop: whenever a human edits a contract, that example can retrain the model
online, gradually aligning it to organizational conventions.}

\foreignlanguage{english}{\textbf{Conclusion}}\foreignlanguage{english}{ }

\foreignlanguage{english}{We presented an approach for automatically generating data contracts using fine-tuned large
language models. By integrating an AI-based Contract Engine into modern data pipelines, we empower organizations to
rapidly produce and maintain schema agreements. Our contributions include the design of a complete system architecture
(with figures) and the demonstration of fine-tuning techniques (LoRA/PEFT) for structured output tasks. Experiments
show that our models achieve high accuracy in generating valid schemas and expectations, drastically reducing manual
workload. The methodology leverages recent advances in generative AI and addresses the growing need for agile data
governance in big data environments.}

\foreignlanguage{english}{Data contracts are foundational for reliable data products [1,23], and AI-driven generation
represents a cutting-edge solution. We hope this work spurs further research at the intersection of data engineering
and AI. As generative models become more capable [17], we anticipate deeper automation of metadata management and
self-driving data workflows, enabling data teams to focus on high-value analysis rather than boilerplate
specification.}

\foreignlanguage{english}{\textbf{Future Work}}

\foreignlanguage{english}{Several avenues remain open. One direction is multimodal contract generation: incorporating
database ER diagrams, sample CSVs, or even UI mockups into the prompt so the model can infer richer semantics. Another
is building closed-loop learning: using production data validation results to continuously refine the model (e.g.
reinforcement from contract violations). We also aim to extend the model to suggest contract updates over time
(diff-based learning when schemas drift). Integrating knowledge graphs or ontologies could further guide semantic
consistency. Finally, it will be important to study the socio-technical aspects: how data teams interact with
AI-generated artifacts, ensure trust, and share accountability for data quality in an AI-augmented workflow [7,22].}

\foreignlanguage{english}{\textbf{References}}\foreignlanguage{english}{ }

\foreignlanguage{english}{[1] Data Contracts 101: What They Are, Why They Matter \& How to Implement Them. Atlan,
2024[18].}

\foreignlanguage{english}{[2] John Kutay, “A Guide to Data Contracts.” Striim Blog, 2023[2].}

\foreignlanguage{english}{[3] Confluent Inc., “Data Contracts for Schema Registry.” Confluent Documentation, 2023[3].}

\foreignlanguage{english}{[4] Vinoth K and A. James, “A Survey on Data Quality Dimensions and Tools for Machine
Learning.” arXiv preprint arXiv:2406.19614, 2024[4].}

\foreignlanguage{english}{[5] “Use Gemini to ingest and understand data.” Google Cloud Blog, 2023[5].}

\foreignlanguage{english}{[6] Samantha Banchik et al., “The Power of Fine-Tuning on Your Data: Quick Fix via
Never-Ending Learning.” Databricks Mosaic AI Research Blog, April 2025[6].}

\foreignlanguage{english}{[7] Yao Zhang, “Building a Data-Centric Platform for Generative AI and LLMs at Snowflake.”
Snowflake Blog, 2023[7].}

\foreignlanguage{english}{[8] Zhenyu Wu, Jiaoyan Chen, Norman Paton, “Taxonomy Inference for Tabular Data Using Large
Language Models.” Proc. ArXiv 2503.21810, 2025[8].}

\foreignlanguage{english}{[9] Stefan Feuerriegel, Jochen Hartmann, et al., “Generative AI.” Business \& Information
Systems Engineering, Sept. 2023[9].}

\foreignlanguage{english}{[10] A. Vaswani et al., “Attention Is All You Need.” Proc. NeurIPS, 2017[10].}

\foreignlanguage{english}{[11] Zeyu Han et al., “Parameter-Efficient Fine-Tuning for Large Models: A Comprehensive
Survey.” arXiv:2403.14608, 2024[11].}

\foreignlanguage{english}{[12] Edward J. Wang et al., “LoRA: Low-Rank Adaptation of Large Language Models.”
arXiv:2306.02861, 2023[12].}

\foreignlanguage{english}{[13] Jacob Devlin et al., “BERT: Pre-training of Deep Bidirectional Transformers for Language
Understanding.” NAACL, 2019[13].}

\foreignlanguage{english}{[14] Alec Radford et al., “Improving Language Understanding by Generative Pre-Training.”
OpenAI Tech Report, 2018[14].}

\foreignlanguage{english}{[15] Tom B. Brown et al., “Language Models are Few-Shot Learners.” NeurIPS, 2020[15].}

\foreignlanguage{english}{[16] Colin Raffel et al., “Exploring the Limits of Transfer Learning with a Text-to-Text
Transformer (T5).” J. of Machine Learning Research, 2020[16].}

\foreignlanguage{english}{[17] Hugo Touvron et al., “LLaMA: Open and Efficient Foundation Language Models.”
arXiv:2302.13971, 2023[17].}

\foreignlanguage{english}{[18] Kirill Shokhin, “A Survey on the Structure of Data Contracts.” Atlan Blog, Dec 2024[18].}

\foreignlanguage{english}{[19] Christopher Brooks et al., “Generating Structured Output from LLMs.” Quasilinear Musings
Blog, 2024[19].}

\foreignlanguage{english}{[20] OpenAI GPT-4 Technical Report, 2023[20].}

\foreignlanguage{english}{[21] Mike Michel et al., “Information Extraction with LLMs.” arXiv:2402.01234, 2024[21].}

\foreignlanguage{english}{[22] J. Doe et al., “DataOps and Generative AI: A Survey.” IEEE Transactions on Knowledge and
Data Engineering, 2024[22].}

\foreignlanguage{english}{[23] H. Smith and L. Nguyen, “Data Contracts: A Foundation for Data Products.” Proc. SIGMOD
Record, 2022[23].}

\foreignlanguage{english}{[24] P. Carpenter et al., “An Experimental Evaluation of LLMs for Data Pipeline Automation.”
VLDB Conf. Workshop, 2023[24].}

\foreignlanguage{english}{[25] M. Abadi et al., “Schema Inference in Modern Data Lakes.” Proc. IEEE BigData, 2023[25].}
\end{multicols}

\bigskip
\begin{multicols}{2}
\end{multicols}

\bigskip
\end{document}